\documentclass[a4paper,11pt]{article}
\pdfoutput=1  
\usepackage[utf8]{inputenc}
\usepackage{jheppub}
\usepackage{amssymb}
\usepackage{physics}
\usepackage{graphicx}
\usepackage{hyperref}
\usepackage{slashed}
\usepackage{orcidlink}
\usepackage{bbm}

\newcommand{\be}{\begin{equation}}
\newcommand{\ee}{\end{equation}}
\newcommand{\bea}{\begin{eqnarray}}
\newcommand{\eea}{\end{eqnarray}}

\DeclareMathOperator{\Eig}{Eig}

\usepackage{color}
\definecolor{colormp}{RGB}{0,130,10}

\begin{document}
 \title{
  Towards a Higgs mass determination in asymptotically safe gravity with a dark portal
 }
 \author[a]{Astrid Eichhorn\orcidlink{0000-0003-4458-1495},}
 \author[a,b]{Martin Pauly\orcidlink{0000-0002-7737-6204},}
 \author[c]{Shouryya Ray\orcidlink{0000-0003-4754-0955}}

\affiliation[a]{CP3-Origins,  University  of  Southern  Denmark,  Campusvej  55,  DK-5230  Odense  M,  Denmark}
\affiliation[b]{Institut f\"ur Theoretische Physik, Ruprecht-Karls-Universit\"at Heidelberg, \\
 Philosophenweg 16, 69120 Heidelberg, Germany}
\affiliation[c]{Institut f\"ur Theoretische Physik and W\"urzburg-Dresden Cluster of Excellence ct.qmat, TU Dresden, 01062 Dresden, Germany}

\emailAdd{eichhorn@cp3.sdu.dk}
\emailAdd{m.pauly@thphys.uni-heidelberg.de}
\emailAdd{shouryya.ray@tu-dresden.de}

\abstract{
There are indications that an asymptotically safe UV completion of the Standard Model with gravity could constrain the Higgs self-coupling, resulting in a prediction of the Higgs mass close to the vacuum stability bound in the Standard Model. 
The predicted value depends on the top quark mass and comes out somewhat higher than the experimental value if the current central value for the top quark mass is assumed. Beyond the Standard Model, the predicted value also depends on dark fields coupled through a Higgs portal.
Here we study the Higgs self-coupling in a toy model of the Standard Model with quantum gravity that we extend by a dark scalar and fermion. 
Within the approximations used in \cite{Eichhorn:2020kca}, there is a single free parameter in the asymptotically safe dark sector, as a function of which the predicted (toy model) Higgs mass can be lowered due to mixing effects if the dark sector undergoes spontaneous symmetry breaking.
}

\maketitle

\section{Introduction}
The discovery of the Higgs particle \cite{ATLAS:2012yve,CMS:2012qbp} with a mass of approximately $m_H \approx 125\, \text{GeV}$ has striking implications for the Higgs potential at large field values.
If the Higgs particle were only slightly heavier, a Landau pole well before the Planck scale would signal the onset of strong coupling or the presence of new degrees of freedom below the Planck scale $M_\text{Pl}$ \cite{Cabibbo:1979ay,Dashen:1983ts,Lindner:1985uk}. 
Similarly, if the Higgs particle was only slightly lighter, vacuum stability considerations would signal the presence of new physics below the Planck scale \cite{Arnold:1989cb,Sher:1993mf,Sher:1988mj,Arnold:1991cv,Espinosa:1995se,Casas:1994qy,Casas:1996aq,Ellis:2009tp,Elias-Miro:2011sqh,Bezrukov:2012sa,Degrassi:2012ry,Buttazzo:2013uya,Espinosa:2015qea,Bednyakov:2015sca,Espinosa:2016nld}.

For the measured mass of the Higgs boson, the Standard Model (SM) remains self-consistent without new physics below the Planck scale. The electroweak potential might exhibit more than one local minimum. Depending on the depth of the additional minima, the SM vacuum at $246\,\text{GeV}$ is either stable or metastable. Its stability is highly sensitive to the top quark mass \cite{Bezrukov:2014ina}. For a top quark mass of $172.8\,\text{GeV}$ \cite{ParticleDataGroup:2020ssz}, the electroweak vacuum is metastable, with a life-time exceeding the age of the universe in the absence of higher-order terms in the microscopic Higgs potential \cite{Branchina:2013jra,Gies:2013fua,Branchina:2014usa,Gies:2014xha,Eichhorn:2015kea,Borchardt:2016xju}.  However, the top quark mass is not yet precisely determined. The precise relation between the experimentally measured and the theoretically relevant parameter remains uncertain \cite{Hoang:2020iah}. The fate of the SM vacuum hence remains an open question.

A different perspective on the question of stability is provided by the map between the value of the quartic coupling at microscopic scales and the Higgs mass. A vanishing value of the quartic coupling at the Planck scale maps to a Higgs mass of $129\,\text{GeV}$ \cite{Bezrukov:2012sa} at a top pole mass of $173\,\text{GeV}$ at three loops and under the assumption that no higher-order  interactions are present. Correspondingly, a slightly negative value of the quartic coupling maps to a Higgs mass of $125\,\text{GeV}$ (at fixed top mass). Such a map between Planck-scale physics and electroweak scale physics is an appropriate point of view when ultraviolet (UV) completions of the SM are explored. A UV completion which includes gravity may determine the value of the Higgs quartic coupling at the Planck scale. Its value maps to electroweak scales, and results in a prediction of the Higgs mass. The corresponding effective potential for the Higgs may feature a single minimum at $246\,\text{GeV}$ (vacuum stability), or an additional, global minimum at higher scales (vacuum metastability).

At the same time, the Renormalization Group (RG) flow of the Higgs quartic coupling is highly sensitive to new degrees of freedom below the Planck scale. Such degrees of freedom may for instance constitute a dark matter component in the Universe's energy budget. Within any such extensions of the SM one can ask:
Do the new degrees of freedom render the electroweak vacuum stable or do they destabilize the electroweak vacuum further? 

Many extensions of the SM do not conclusively answer this question. Implemented as an effective field theory (EFT), the values of couplings correspond to free parameters. The resulting parameter space contains a wide range of possible phenomenological consequences. To improve predictivity in these situations, an additional symmetry principle may be imposed. Scale symmetry, both classical and quantum, is a strong contender for this additional symmetry principle, see, e.g., Ref.~\cite{Meissner:2006zh,Foot:2007as,Shaposhnikov:2008xi,Iso:2009ss,Shaposhnikov:2009pv,Alexander-Nunneley:2010tyr,Carone:2013wla,Englert:2013gz,Farzinnia:2013pga,Holthausen:2013ota,Antipin:2013exa,Hill:2014mqa,Oda:2015gna,Helmboldt:2016mpi,Shaposhnikov:2018jag,Prokopec:2018tnq,Shaposhnikov:2018nnm,Ferreira:2018itt,Jung:2019dog,Wetterich:2019qzx,Braathen:2020vwo,Eichhorn:2020sbo} and references therein for various aspects of classically or quantum scale invariant (extended) Higgs sectors.

Asymptotically safe quantum gravity implements quantum scale symmetry at high energies. Just as for any other symmetry, deformations of the symmetry can only occur along the relevant directions of the corresponding RG fixed point. For classical symmetries, these relevant directions follow canonical power counting; for instance a mass parameter is the only relevant deformation of a shift-symmetric scalar field theory in four dimensions. In contrast, for quantum scale symmetry, the relevant directions do not follow canonical power counting, because quantum effects correct the scaling exponents. Therefore, canonically relevant or marginal couplings may be irrelevant and thus be predicted \emph{at all scales}, even in the deep infrared (IR), where quantum scale symmetry is no longer realized. 
 Thereby, a realization of asymptotic safety or quantum scale symmetry \cite{Wetterich:2019qzx} can enhance predictivity over the EFT setting, both for the study of dark matter \cite{Eichhorn:2017als,Reichert:2019car,Hamada:2020vnf,Eichhorn:2020kca}, as well as the Higgs mass \cite{Shaposhnikov:2009pv} and further SM couplings \cite{Harst:2011zx,Eichhorn:2017ylw,Eichhorn:2017lry,Eichhorn:2018whv,Alkofer:2020vtb}.

 In this paper, we leverage the enhanced predictivity in an extension of the SM.
We consider a toy model for the SM, coupled to a dark sector that contains a dark matter candidate. We study the impact of the dark sector on the Higgs mass. To do so we introduce the asymptotic safety paradigm in more detail in Sec.~\ref{sec:as_qg}. In Sec.~\ref{sec:higgs_mass_and_rg} we discuss the general relation between the UV and the IR Higgs mass, before we go into detail on a specific model in Sec.~\ref{sec:higgs_mass_extended}. In Sec.~\ref{sec:conclusion} we present our conclusions.

\section{Asymptotically Safe Quantum Gravity}
\label{sec:as_qg}
Quantum fluctuations render couplings in a quantum field theory dependent on the energy scale. In the Wilsonian functional RG approach to renormalization one integrates out these fluctuations momentum-shell-wise from high to low energy scales.
The resulting dependence on the energy scale $k$ is encoded in the beta functions $\beta_g = k \partial_k g$ for a dimensionless coupling $g = \bar{g} k^{-d_g}$. Here $d_g$ is the canonical mass dimension of the dimensionful coupling $\bar{g}$. 

\subsection{Fixed points and predictivity}
Quantum scale invariance is realized by a fixed point of all beta functions in a given theory, i.e., $\beta_{g_i}=0\, \forall i$ at $g_i=g_{i\,\ast}$. At such a fixed point, all couplings become independent of the energy scale, $\beta_{g_i} = 0$. As a result the theory does not feature an intrinsic cutoff scale at which it breaks down. It therefore constitutes a candidate for a fundamental microscopic description. 
Most importantly a fixed point generates predictivity by dividing the couplings into relevant and irrelevant ones:
If the couplings initially deviate from their fixed-point values in the UV, such deviations can either grow or shrink towards the IR. This is encoded in the critical exponents $\theta_i$, i.e., the eigenvalues of the stability matrix multiplied by an additional negative sign
\be
  \theta_i = - \Eig \pdv{\beta_{g_j}}{g_k}\, \Big|_{g_j=g_{j\, \ast}}
.\ee 
A positive (real part of a) critical exponent corresponds to a relevant direction; deviations from quantum scale symmetry in the corresponding eigendirection grow towards the IR. This implies that a range of IR values is compatible with quantum scale symmetry in the UV and experimental input is needed to determine the IR value. Conversely, a negative (real part of a) critical exponent corresponds to an irrelevant direction; deviations from quantum scale symmetry in the corresponding eigendirection shrink towards the IR. This implies that there is just one single IR-value for the coupling that complies with quantum scale symmetry in the UV and a prediction from theory can be compared to experiment.\\
 In a setting where interactions are near-perturbative, each eigendirection is roughly aligned with a corresponding coupling. If we classify a coupling as (ir-)relevant in the following, we mean that there exists an eigendirection with huge overlap with the corresponding coupling that is (ir-)relevant. 

Each relevant direction requires a measurement. Once all relevant couplings are fixed in this way, the theory becomes predictive. A finite number of relevant directions guarantees predictivity.
 In contrast, an EFT setting for the same field content and symmetries contains infinitely many free parameters, as the couplings of all operators need to be measured. Fixing infinitely many of these parameters by requiring asymptotic safety
 makes it possible to overcome the problems of perturbative quantum gravity and renders a quantum theory of the metric predictive.

As an example of scale symmetry, consider a non-interacting fixed point. It realizes scale symmetry in a simple way, by switching off the effect of all quantum fluctuations.
The fixed points' critical exponents are given by the canonical mass dimensions of the corresponding dimensionful couplings. In a (quasi-)local Lagrangian containing all terms compatible with a particular symmetry, this renders infinitely many couplings irrelevant and a finite set relevant. Once all relevant couplings are measured, the theory becomes predictive. QCD is a paradigmatic example for this scenario. Asymptotically safe gravity is the generalization of this scenario to an interacting fixed point for quantum gravity.

\subsection{The asymptotic safety scenario for quantum gravity and matter}\label{sec:ASQGM}
If the asymptotic safety scenario for quantum gravity is realized, the QFT for the metric and the SM can be extended to arbitrarily high energies due to the presence of an interacting fixed point, see \cite{Eichhorn:2017egq,Eichhorn:2018yfc,Pawlowski:2020qer} for recent reviews, \cite{Percacci:2017fkn,Reuter:2019byg,Eichhorn:2020mte,Reichert:2020mja} for pedagogical introductions, and \cite{Donoghue:2019clr,Bonanno:2020bil} for critical discussions of the current status of the field. Based on the methodology pioneered in \cite{Reuter:1996cp}, evidence for the existence of such a fixed point has been accumulated in pure gravity using functional RG techniques\cite{Falls:2013bv,Falls:2014tra,Falls:2017lst,Falls:2018ylp,Kluth:2020bdv,Benedetti:2009rx,Gies:2016con,Falls:2020qhj,Knorr:2021slg,Donkin:2012ud,Eichhorn:2018akn,Christiansen:2017bsy,Denz:2016qks,Christiansen:2015rva,Christiansen:2012rx,Christiansen:2014raa,Christiansen:2016sjn,Knorr:2017fus}. 
In addition, lattice techniques \cite{Ambjorn:2012jv,Loll:2019rdj,Laiho:2016nlp,Bassler:2021pzt} and analytical tensor-model techniques \cite{Eichhorn:2018phj}
are used to search for asymptotic safety.
A large body of literature indicates the existence of a fixed point with a finite (and small -- typically two to three) number of relevant directions.  \\ 
The fixed point persists under the inclusion of certain sets of matter fields. In particular, there are indications that it supports SM-like matter \cite{Dona:2013qba,Meibohm:2015twa,Christiansen:2017cxa,Alkofer:2018fxj,Wetterich:2019zdo}. The resulting combined fixed point has interesting properties, both, in the gravity and the matter sector. In the gravity sector it is shifted with respect to the one without matter: it appears at different fixed-point values for the gravitational couplings.\\
The relevant directions in the gravity sector include the Newton coupling and the cosmological constant, making the UV fixed point compatible with the measured IR values. Higher order curvature terms inevitably appear at the fixed point. Their couplings are canonically irrelevant starting from terms cubic in the curvature, while the terms quadratic in the curvature are canonically marginal. Various results studying extended truncations in the gravity sector \cite{Falls:2013bv,Falls:2014tra,Falls:2017lst,Falls:2018ylp,Kluth:2020bdv,Falls:2020qhj,Knorr:2021slg} and symmetry identities at the fixed point \cite{Eichhorn:2018akn,Eichhorn:2018ydy,Eichhorn:2018nda} as well as the existence of a weak-gravity bound in gravity matter systems \cite{Eichhorn:2016esv,Christiansen:2017gtg, Eichhorn:2017eht, Eichhorn:2019yzm,deBrito:2021pyi} suggest that the gravitational fixed point could be near-perturbative in nature: canonically irrelevant couplings typically remain irrelevant at the interacting gravity-matter fixed point because the scaling spectrum is close to canonical scaling. This implies a finite number of relevant directions and hence predictivity. 
  
In the matter sector, it is impossible to set all interactions to zero, because the interacting nature of the gravitational fixed point necessarily percolates into the matter sector. Nevertheless, there may be a choice of distinct universality classes, depending on the gravitational fixed-point values:\\
 At the maximally symmetric fixed point \cite{Eichhorn:2017eht}, higher-order interactions are necessarily present \cite{Eichhorn:2011pc, Eichhorn:2012va}, but all canonically marginal and relevant couplings of the SM may be set to zero. At this fixed point, the Higgs quartic coupling is irrelevant, resulting in a prediction of the Higgs mass \cite{Shaposhnikov:2009pv}. All gauge couplings may become asymptotically free at this fixed point \cite{Daum:2009dn,Daum:2010bc,Folkerts:2011jz,Christiansen:2017cxa} and Yukawa couplings may be relevant for an appropriate range of the gravitational couplings \cite{Oda:2015sma,Eichhorn:2016esv, Eichhorn:2017eht}.\\
At a second fixed point, where shift symmetry in the scalar sector is explicitly broken, higher-order as well as marginally relevant interactions are non-vanishing. At this fixed point, some Yukawa couplings may be finite and irrelevant \cite{Eichhorn:2017ylw,Eichhorn:2018whv,Alkofer:2020vtb}, the Abelian gauge coupling may be irrelevant \cite{Harst:2011zx,Eichhorn:2017lry} and the quartic Higgs  and non-minimal Higgs-curvature couplings may also be irrelevant \cite{Wetterich:2019rsn,Eichhorn:2020sbo}, resulting in very high predictive power of this universality class.

Similarly, a number of Beyond SM settings with gravity have been studied, with some indications for enhanced predictive power from asymptotic safety \cite{DeBrito:2019rrh,Reichert:2019car,Eichhorn:2020kca,Kowalska:2020gie,Kowalska:2020zve,Hamada:2020vnf}.

\subsection{Functional Renormalization Group}
The results reviewed above are obtained with functional RG (FRG) techniques that we also rely on in this paper.
The FRG implements the idea of momentum-shell wise integration of quantum fluctuations in the path integral. The method generalizes the effective action to the scale-dependent effective action $\Gamma_k$. $\Gamma_k$ is the effective action in a theory with an IR cutoff at $k$ and is obtained by integrating out all quantum fluctuations with momentum $p^2>k^2$, while supressing those with momenta $p^2<k^2$.
This is implemented by introducing a regulator function $R_k(p^2)$. The regulator acts as a mass for fluctuations with momenta smaller than $k$. By integrating out fluctuations with larger momenta one obtains the flow equation \cite{Wetterich:1992yh,Ellwanger:1993mw,Morris:1993qb}
\be
  \label{eq:flow_equation}
  k \partial_k \Gamma_k = \frac{1}{2} \operatorname{STr}
  \left( \frac{k \partial_k R_k}{\Gamma^{(2)}_k+R_k} \right)
.\ee
Here, $\Gamma_k^{(2)}$ is the second variation with respect to the corresponding fluctuation fields. The supertrace is over all internal and spacetime indices, as well as over momentum space with an additional negative sign for fermions. When integrating out all fluctuations, one recovers the standard effective action $\Gamma_{k \to 0} = \Gamma$.

Eq.~\eqref{eq:flow_equation} is formally exact and describes how $\Gamma_k$ evolves upon integrating out quantum fluctuations. In practical applications, its exact character is lost due to the necessity to truncate the space of couplings:
Quantum fluctuations generically generate all operators compatible with a particular symmetry. This entails that $\Gamma_k$ typically consists of an infinite tower of operators compatible with the symmetries under consideration. In practical computations, one has to restrict to a particular set of operators, i.e., to a truncation. This introduces a systematic uncertainty in the evaluation of Eq.~\eqref{eq:flow_equation} that can be controlled by the choice of an appropriate truncation scheme. For reviews of the method, see Refs.~\cite{Pawlowski:2005xe,Gies:2006wv,Delamotte:2007pf,Rosten:2010vm,Dupuis:2020fhh}.

When one applies Eq.~\eqref{eq:flow_equation} to gravitational systems, additional subtleties arise. 
Most importantly, the flow equation is set up in Euclidean signature, requiring an analytical continuation in QFTs on a flat background. In the presence of a fluctuating background, an analogous analytical continuation is not possible \cite{Baldazzi:2018mtl}. 
Thus, results on asymptotically safe gravity matter systems are obtained in a Euclidean setting; see  \cite{Manrique:2011jc} for a study that singles out a time direction and \cite{Bonanno:2021squ} for the analytic continuation of the graviton propagator about a flat background as well as \cite{Draper:2020bop,Draper:2020knh,Platania:2020knd,Knorr:2021niv} for discussions of the analyticity properties of this propagator in asymptotic safety.

\section{Review: Higgs mass and RG flow in the SM and beyond}
\label{sec:higgs_mass_and_rg}
To set the stage for our analysis, we review the Higgs mass in the SM, an extension of the SM by a portal to a dark scalar as well as the (conjectured) asymptotically safe versions of both for which promising indications have been found in the literature, see Sec.~\ref{sec:ASQGM}.

\subsection{Higgs mass in the SM}
In the SM, the Higgs mass is
\be
 M_\text{Higgs} = \sqrt{2 \lambda_\text{H}(k_\text{IR}) v_\text{Higgs}^2}
,\ee
where $\lambda_\text{H}$ is the Higgs quartic coupling evaluated at an IR scale $k_\text{IR}$ (typically chosen as the top mass scale) and $v_\text{Higgs} \approx 246\, \text{GeV}$ is the vacuum expectation value (vev) of the Higgs field at the electroweak minimum.
The Higgs vev is known from the measured masses of the weak gauge bosons. The Higgs mass measurement fixes the quartic coupling at the electroweak scale. This measurement can be used to draw inferences on the microphysics at larger energies, e.g., the Planck scale. This map between Planck scale and electroweak scale \emph{starts} with couplings specified at the Planck scale, and follows the RG flow in its proper direction to the IR.
Applied to different microscopic models, it allows to identify those that yield the observed Higgs mass.
Within the SM, the measured Higgs mass  is achieved by starting from a near-vanishing value of the quartic coupling at the Planck scale.   In fact, under the assumption of vanishing higher-order couplings, the measured Higgs mass requires a slightly negative Higgs quartic coupling at the Planck scale. \\
This finding depends on the top-mass:
 because the top Yukawa contribution nearly cancels the gauge coupling contribution to the flow of the Higgs quartic coupling, the Higgs quartic coupling changes very little over a large range of scales. This balance is highly sensitive to the top Yukawa coupling and thus small changes in the top mass strongly alter the flow of the quartic coupling \cite{Bezrukov:2014ina}. The top mass measurement is non-trivial, as is its translation into the  parameter that enters the running \cite{Hoang:2020iah}; thus  a significant uncertainty on the top Yukawa coupling remains.
 Novel direct measurements of the top pole mass \cite{CMS:2017iqf,CMS:2019esx,ATLAS:2019guf} point to a slightly lower value than the central value recorded in \cite{ParticleDataGroup:2020ssz}.
This difference of just 1--2\,GeV in the top quark mass is sufficient to shift the Higgs quartic coupling at the Planck scale from a negative value to zero while keeping its IR value fixed to the measured one.

Using the running quartic coupling to RG improve the electroweak potential results in a potential on the boundary of stability and metastability. A metastable electroweak vacuum, tied to the larger top mass,  has a lifetime longer than the age of the universe \cite{Bezrukov:2012sa,Degrassi:2012ry,Buttazzo:2013uya,Bezrukov:2014ina,Elias-Miro:2011sqh}. Higher-order couplings can change this conclusion \cite{Branchina:2013jra,Gies:2013fua,Branchina:2014usa,Branchina:2014rva,Gies:2014xha,Eichhorn:2015kea,Borchardt:2016xju}.

Going beyond the SM, one defines a model at some microscopic scale $\Lambda$, e.g., the Planck scale. Any physically viable model must feature a potential that is bounded from below at this scale $\Lambda$. Following the RG flow towards lower scales deforms that potential at field values $\phi \lesssim \Lambda$ and can therefore not result in an unstable electroweak potential. The minimum quartic coupling that is achievable in such a setting translates into a lower bound on the Higgs mass.
 The lower bound on the Higgs mass is a prediction of the particular model.
 For instance, within the SM defined at $\Lambda= M_{\rm Pl}$, a minimum quartic coupling $\lambda_{\rm H}(M_{\rm Pl})=0$ results in a Higgs mass of $133\,\text{GeV}$ (at one loop, $128\,\text{GeV}$ at two loop) \footnote{These values were obtained with a top mass $M_t=172.8 \text{GeV}$ \cite{ParticleDataGroup:2020ssz}.} 
 
An asymptotically safe UV completion with gravity  extends the theoretical validity of the quantum field theoretic description to infinite energies, $\Lambda \to \infty$ \footnote{If a different UV completion is valid above the scale $\Lambda \gg M_\text{Pl}$, but below $\Lambda$ effective metric degrees of freedom govern the RG flow, then \emph{effective asymptotic safety} could play a role and enhance predictivity, see e.g.,~\cite{deAlwis:2019aud,Held:2020kze,Basile:2020dzh,Basile:2021krk,Basile:2021krr}.}.
In this scenario a UV fixed point (i) provides a UV completion, (ii) could fix some of the marginal couplings in the  gauge-Yukawa sector and (iii) is expected to predict the Planck-scale value of the Higgs quartic coupling.
This determines the Higgs mass in an asymptotically safe model as a function of the relevant couplings of that model.

 In the following, we first review the Higgs mass in a portal model to dark scalars in Sec.~\ref{sec:Higgsmassportal} and the Higgs mass in the (conjectured) asymptotically safe SM in Sec.~\ref{Sec:HiggsmassAS}, before combining the two and reviewing the status of asymptotically safe portal models in Sec.~\ref{sec:HPandAS}. This provides the basis for our new results on Higgs mass bounds in asymptotically safe portal models.

\subsection{Higgs mass bounds in bosonic portal models}
\label{sec:Higgsmassportal}
A gauge singlet $\phi_d$ may constitute dark matter \cite{Silveira:1985rk,McDonald:1993ex,Bento:2000ah,Burgess:2000yq,Bento:2001yk,McDonald:2001vt,Davoudiasl:2004be,
Patt:2006fw,OConnell:2006rsp,Barger:2007im,He:2007tt,Barger:2008jx,He:2008qm} and lower the lower bound on the Higgs mass \cite{Gonderinger:2009jp,Clark:2009dc,Lerner:2009xg,Gonderinger:2012rd,Chen:2012faa,Cline:2013gha,Khoze:2014xha,Eichhorn:2014qka}.
The dark scalar $\phi_d$ obeys a discrete $\phi_d \rightarrow - \phi_d$ symmetry to ensure its stability and is coupled to the Higgs scalar $\Phi$ via a portal operator 
\be
\mathcal{L}_\text{HP} = \frac{\lambda_\text{P}}{4} \Phi^\dagger \Phi \phi_d^2\label{eq:portal}
.\ee
The portal coupling enables thermal production, see \cite{Roszkowski:2017nbc,Arcadi:2017kky} for reviews on thermally produced dark matter and \cite{Arcadi:2019lka} for a review of portal dark matter. Through the portal, dark bosonic fluctuations change the flow of the Higgs quartic coupling, counteracting the effect of top quark fluctuations and lowering the Higgs mass. Direct and indirect observational bounds on the value of the portal coupling and the value of the dark scalar mass are reported in \cite{GAMBIT:2017gge,GAMBIT:2018eea}.

The portal coupling has two distinct effects on the Higgs mass:\\
First, starting from a fixed value of the Higgs quartic coupling in the UV, the additional bosonic fluctuations lower the quartic coupling in the IR. This is because the fluctuations of $\phi_d$ contribute a term $c_\text{P} \lambda_\text{P}^2$ with $c_\text{P}>0$ to $\beta_\lambda$. The sign of $\lambda_\text{P}$  is irrelevant for this effect, which has been discussed in, e.g.,\cite{Gonderinger:2009jp,Cline:2013gha,Khoze:2014xha} and with the FRG\footnote{For an FRG study of vacuum stability in the Higgs portal to fermionic dark matter, see \cite{Held:2018cxd}.} in \cite{Eichhorn:2014qka}.\\
Second, if the dark scalar also acquires a vacuum expectation value, the resulting mixing between dark and visible scalar will affect the measured Higgs mass at tree level through off-diagonal terms in the scalar mass matrix. 
The physical masses are the eigenvalues of the mass matrix. Due to the mixing between $\Phi$ and $\phi_d$, the corresponding eigenvalues repel each other, as we demonstrate for a  potential with  real dark scalar coupled to the radial mode of the Higgs, $\phi_v$ of the form 
\be
V(\phi_v, \phi_d)= \frac{\bar{m}_v^2}{2} \phi_v^2 + \frac{\lambda_v}{8}\phi_v^4 + \frac{\lambda_{\rm HP}}{4}\phi_v^2\phi_d^2 + \frac{\bar{m}_d^2}{2} \phi_d^2 + \frac{\lambda_d}{8}\phi_d^4,
\ee
that we will investigate in Sec.~\ref{sec:higgs_mass_extended}.  In the symmetry-broken regime
we rewrite the potential in terms of the 
vacuum expectation values $\expval{\phi_{v(d)}} = v_{v(d)}$
\be
  v_{v(d)}^2 = 2 \frac{m^2_{v(d)} \lambda_{d(v)}- m^2_{d(v)} \lambda_\text{HP}}{\lambda_\text{HP}^2 - \lambda_v \lambda_d}
.\ee
 The potential can then be written as
\be
\label{eq:potential}
  V(\phi_v, \phi_d) = \frac{\lambda_v}{8} \left(\phi_v^2 - v_v^2 \right)^2 + \frac{\lambda_d}{8} \left(\phi_d^2 - v_d^2 \right)^2 + \frac{\lambda_\text{HP}}{4} \left(\phi_v^2 - v_v^2 \right) \left( \phi_d^2 - v_d^2 \right)
,\ee
 where we neglected a constant shift in the potential that is irrelevant in the absence of gravity. 
In this potential, fluctuations around the minimum have mass
\be
  \label{eq:masses}
  M^2_{v/d} = \frac{1}{2} \left( \lambda_v v_v^2 + \lambda_d v_d^2 \pm \sqrt{(\lambda_v v_v^2 - \lambda_d v_d^2)^2 + 4 \lambda_\text{HP} v_v^2 v_d^2} \right)
.\ee

The repulsion between the two eigenvalues lowers the smaller of the two masses and increases the larger one.
When the dark scalar is heavier than the Higgs this leads to a decrease of the Higgs mass \cite{Lebedev:2012zw,Elias-Miro:2012eoi,Falkowski:2015iwa}. 

This effect does not vanish even for very large vacuum expectation values of the dark scalar.
To describe the effect of the dark scalar in this limit, one solves the equation of motion for $\phi_d$ in the limit of slowly varying field in the potential \eqref{eq:potential}. By reinserting the result into \eqref{eq:potential} one obtains a corrected quartic coupling $\tilde{\lambda}_\text{H}$,
\be
  \tilde{\lambda}_\text{H} = \lambda_\text{H}-\frac{\lambda_\text{HP}^2}{\lambda_d}.
\ee
The correction $\frac{\lambda_\text{HP}^2}{\lambda_d}$ reduces the Higgs quartic coupling and lowers the Higgs mass. It remains finite for large $v_d$. 

At a first glance, the correction appears to violate decoupling theorems in EFT. However, a massive degree of freedom only decouples in the limit of infinite mass, if its coupling to the remaining degrees of freedom is held constant. In the present case, both the dark scalar's mass  and its coupling to the Higgs increase with $v_d^2$, such that a contribution from $\phi_d$ to the effective action remains in the limit $v_d \rightarrow \infty.$

The dark scalar $\phi_d$ might play the role of a dark matter candidate. This is only viable if the discrete $\mathbb{Z}_2$ symmetry for $\phi_d$ is unbroken, otherwise the dark matter decays. Observationally, this option is only viable for a narrow window of masses close to $M_\text{Higgs}/2$, as well as for dark scalar masses $M_d \gtrsim 10^{3}\, \text{GeV}$ \cite{GAMBIT:2017gge,Athron:2018ipf}, where the latter region is associated with fairly large portal couplings $\lambda_\text{HP} \sim 1$. For unbroken $\mathbb{Z}_2$ symmetry, only the first of the two discussed effects contributes to the Higgs mass. This contribution to the flow of $\lambda_\text{H}$ can lower the Higgs mass enough to match observations while maintaining a stable electroweak vacuum, see \cite{Athron:2018ipf}.

\subsection{Higgs mass in asymptotic safety}\label{Sec:HiggsmassAS}
Asymptotically safe gravity-matter models might predict the ratio of the Higgs mass  to the electroweak scale \cite{Shaposhnikov:2009pv}. 
This prediction relies on the irrelevance of the Higgs quartic coupling at an asymptotically safe matter-gravity fixed point.\footnote{Here, we solely focus on the case where the Higgs mass parameter remains relevant. The mass parameter may become irrelevant, see, e.g.,Ref.~\cite{Wetterich:2016uxm} for sufficiently strong gravity fluctuations. Such strong gravity fluctuations are most likely not compatible with the weak-gravity bound discovered in several asymptotically safe matter-gravity systems \cite{Eichhorn:2016esv,Christiansen:2017gtg,Eichhorn:2017eht,Eichhorn:2019yzm}.} 
At one loop and with the added gravitational contribution, the beta function for the Higgs quartic coupling $\lambda_\text{H}$ reads
\bea
  \label{eqn:beta_lambda}
 \beta_{\lambda_\text{H}}= - f_s \lambda_\text{H} 
   &+& \frac{1}{16\pi^2} \left( - 6 y_t^4 + \frac{3}{8}\left(2 g_2^4 + (g_2 + \frac{5}{3} g_Y^2)^2\right) \right)  \nonumber \\
     &+& \frac{1}{16\pi^2} \lambda_{\rm H} \left(12 y_t^2-9 g_2^2-5 g_Y^2 \right) + \frac{3}{2\pi^2}\lambda_{\rm H}^2
.\eea
Here $f_s$ is a function of the gravitational couplings that is independent of the internal indices of the scalar field. It has been computed in \cite{Narain:2009fy, Percacci:2015wwa,Eichhorn:2017als, Pawlowski:2018ixd,DeBrito:2019gdd,Wetterich:2019zdo,Wetterich:2019rsn,Eichhorn:2020sbo} and encodes the gravitational contributions to the beta function. 
Additional contributions come from the gauge and the Yukawa sector, respectively, where we neglect all but the top quark Yukawa. The Yukawa contributions come with the opposite sign to the gauge contribution due to the fermionic nature of the corresponding fluctuations.

 For $f_s<0$, as has been found in \cite{Narain:2009fy, Percacci:2015wwa,Eichhorn:2017als, Pawlowski:2018ixd,DeBrito:2019gdd,Wetterich:2019zdo,Wetterich:2019rsn,Eichhorn:2020sbo}, there is an IR attractive fixed point in $\lambda_{\rm H}$. It lies at either a vanishing or a non-vanishing value of $\lambda_{\rm H}$, depending on the fixed-point structure in the gauge-Yukawa sector. The non-Abelian gauge couplings remain asymptotically free under the impact of quantum gravity \cite{Daum:2009dn,Folkerts:2011jz,Christiansen:2017cxa}, the Abelian gauge coupling may be either asymptotically free or safe \cite{Harst:2011zx,Eichhorn:2017lry,Christiansen:2017gtg}, with indications for an upper bound on its Planck-scale value \cite{Eichhorn:2017lry}. Similarly, the top Yukawa coupling may be asymptotically free or safe, if conditions on the gravitational fixed-point values are met \cite{Oda:2015sma,Eichhorn:2016esv, Eichhorn:2017eht}.

We first assume that Yukawa and gauge couplings become asymptotically free under the impact of gravity such that they can be set to zero to analyse the fixed point for the quartic coupling. The only fixed point at non-negative values for $\lambda_\text{H}$ is
\be
  \label{eqn:higgs_vanishing_fp}
  \lambda_{\text{H}\ast} = 0, \quad \theta_{\lambda_\text{H}} = f_s < 0,
\ee
The quartic coupling $\lambda_\text{H}$ vanishes and is irrelevant.
Gravitational fluctuations hence dampen any deviation from the vanishing fixed-point value. Thereby the quartic coupling (nearly) vanishes at the Planck scale\footnote{Due to the growth of the other SM couplings to non-zero Planck-scale values from their asymptotically free fixed-point values, the critical value of the Higgs quartic coupling that is reached from its vanishing fixed-point value is actually nonzero, but very small.}. It is regenerated below the Planck scale by the other SM couplings which grow towards the IR from their asymptotically free fixed-point values. The resulting IR value for the quartic coupling corresponds to a Higgs mass in the vicinity of the measured value. In fact, this scenario predicts a Higgs quartic coupling at the stability bound of the SM \cite{Shaposhnikov:2009pv}.

We next consider non-vanishing fixed-point values for Yukawa and Abelian gauge coupling, i.e., a second potential gravity-matter universality class. This results in a positive fixed-point value for $\lambda_\text{H}$, given by
\be
  \label{eq:lambda_interacting_fp}
 \lambda_{\text{H}\, \ast}= \frac{5}{48}g_Y^2-\frac{1}{4} y_t^2 + \frac{\pi^2}{3} f_s + \frac{1}{48}\sqrt{\left(12 y_t^2-5g_Y^2 - 16 \pi^2 f_s \right)^2+576 y_t^4 - 100 g_Y^4}.
\ee
The larger fixed-point value translates into a larger IR value, and hence a larger Higgs mass. 
Thus, this universality class appears to be in tension with observations, unless additional degrees of freedom are added, as we will do below.

\begin{figure}
	\centering
	\begin{tabular}{c c}
	\includegraphics[width=0.45\textwidth]{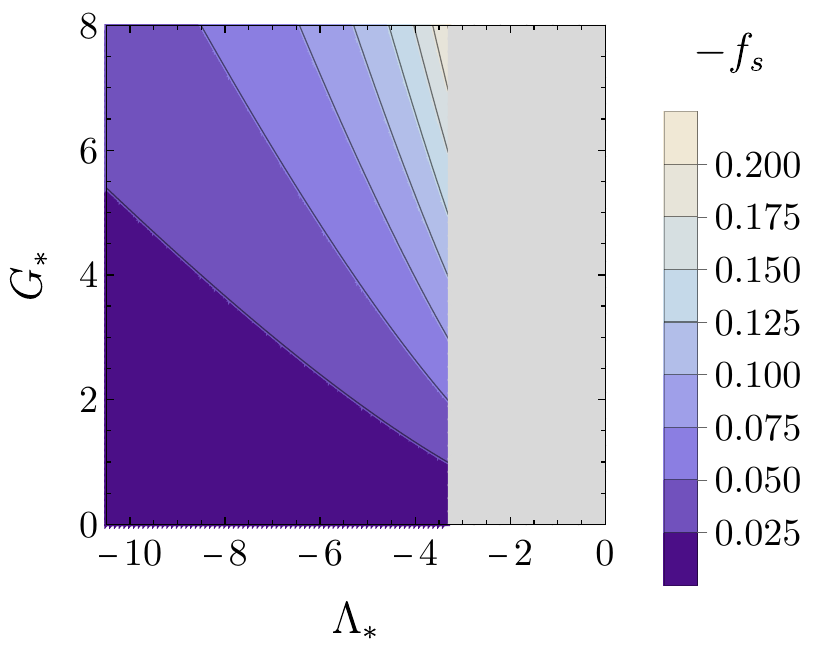}  & \includegraphics[width=0.45\textwidth]{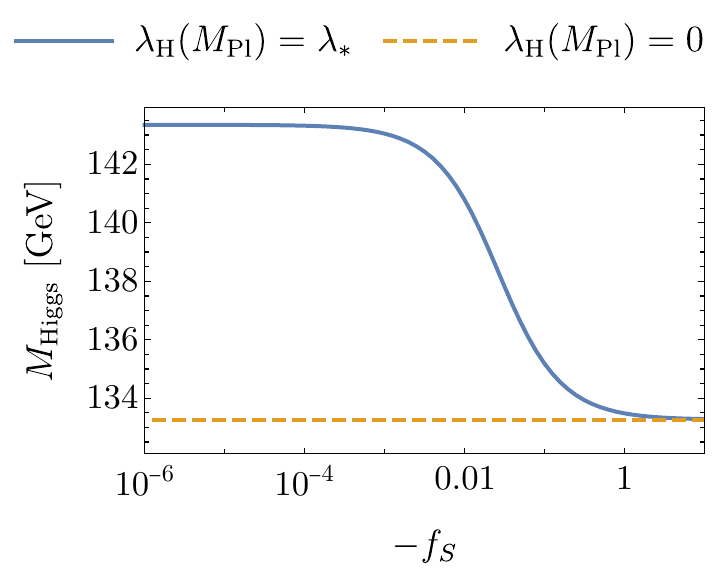}\\
	(a) & (b)
	\end{tabular}\\[1em]
  \includegraphics[width=0.7\textwidth]{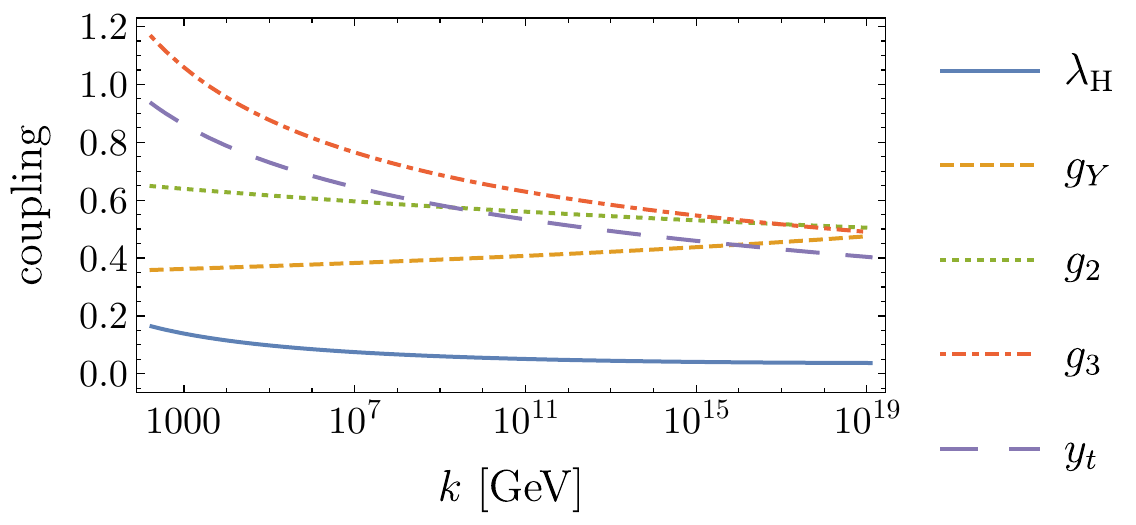}\\
  (c)
    \caption{\label{fig:mass_higgs_sm}
    Here we reverse the RG flow to evolve the IR values for the SM couplings to the scale $M_\text{Pl}$ using the SM one-loop beta functions, implement the fixed point condition \eqref{eq:lambda_interacting_fp} for $\lambda_\text{H}$ and evolve the Higgs quartic coupling to IR scales. (a) Contours of the quantity $f_s$ which encodes the gravitational contributions to $\beta_{\lambda_\text{H}}$ as a function of gravitational couplings $(G,\Lambda)$. The grey region is excluded because it could result in a vanishing top mass, as found within an approximation in \cite{Eichhorn:2017ylw}.
     (b) Higgs mass as a function of $f_s$. For $\abs{f_s} \gg 0$, one approaches the Higgs mass corresponding to the vacuum stability bound with $\lambda_\text{H}(M_\text{Pl}) = 0$ at one loop. (c) Illustration of the flow of SM couplings from the Planck scale down to the scale of the top mass, with the Higgs self-coupling initialized to solve $\beta_{\lambda_\text{H}} = 0$ for $f_s = -0.01$ [cf. Eq.~\eqref{eqn:beta_lambda}] at the Planck scale.}
 \end{figure}
 
 The resulting shift in the Higgs mass depends on the strength of gravitational corrections.
In Fig.~\ref{fig:mass_higgs_sm}, we compute the shift in a one-loop approximation under the assumption that the predictions of Abelian gauge coupling and top quark mass from asymptotic safety match the observed values. To compute the flow, we require the Planck-scale values of the SM couplings as input. We obtain these by reversing the direction of the flow and mapping the IR values $g_Y(k_{\rm IR}) = 0.3587$, $g_2(k_{\rm IR})= 0.6484$, $g_3(k_{\rm IR})= 1.1647$, $y_t(k_{\rm IR})= 0.9080$ at $k_{\rm IR}=172.8\,\text{GeV}$ \footnote{
These values correspond to a top mass of $M_\text{top}=172.8\,\text{GeV}$ and are obtained by three-loop QCD matching and two-loop matching for $y_t$ and $\lambda_\text{H}$ \cite{Bezrukov:2012sa,Chetyrkin:2012rz}.
} to their corresponding Planck-scale values. We use the one-loop beta functions for the top Yukawa coupling and the Abelian hypercharge $g_Y$. Based on this input, we determine $\lambda_{\text{H}\,\ast}$ from \eqref{eq:lambda_interacting_fp} for various values of $f_s$. We assume that $\lambda_{\text{H}\,\ast}$ is the Planck-scale value for the quartic coupling and 
flow to the IR using the SM one-loop beta function and compute the resulting Higgs mass. For $\abs{f_s}\gg 1$ one approaches the limiting case $\lambda_\text{H} \approx 0$. For smaller values of $\abs{f_s}$, the quartic coupling is larger at the Planck scale, translating to higher Higgs masses in the IR.

Assuming the current central value for the top mass, in both scenarios -- i.e., at the Gaussian matter fixed point  and at the interacting matter fixed point with dominant Yukawa contributions -- the resulting IR Higgs mass is (slightly) too large. We take this as a motivation to explore which extensions of the SM could lower the IR Higgs mass and at the same time are compatible with asymptotically safe quantum gravity. The question of Higgs vacuum stability in asymptotic safety has previously been investigated in the context of a model with neutrino masses \cite{Domenech:2020yjf}.

\subsection{Higgs Portal and Asymptotic Safety}\label{sec:HPandAS}
In the previous subsections we have reviewed results on the Higgs mass in the SM and its asymptotically safe extension with gravity. In the SM, requiring vacuum stability at a top mass of $173\,\text{GeV}$ leads to a Higgs mass a few GeV above the measured values. In the asymptotically safe SM, two universality classes could be available for a UV completion. Both predict the Higgs quartic coupling and as a result the Higgs mass. The predictions lie at least a few GeV above the measured value. In the non-gravitational setting, a dark scalar coupled to a Higgs portal allows to reach lower values of the Higgs mass. This motivates us to investigate whether the same mechanism may be available in an asymptotically safe extension of the model with gravity.

In the following we review the implications of quantum scale symmetry on the portal coupling \cite{Eichhorn:2017als}. Quantum gravity fluctuations only produce a free fixed point in the portal coupling in Eq.~\eqref{eq:portal}. At this free fixed point, quantum gravity fluctuations render the portal coupling irrelevant. The dark scalar mass also features a vanishing fixed-point value, but remains relevant unless quantum-gravity fluctuations are very strong.
These results hold even if the Higgs quartic coupling takes on a finite fixed-point value, as reviewed in Sec.~\ref{Sec:HiggsmassAS}. 
For this fixed-point structure the portal coupling must vanish at the Planck scale. As no Yukawa or gauge degrees of freedom regenerate the coupling below the Planck scale, the portal coupling vanishes in the IR\footnote{There is the possibility for a tiny portal coupling that is generated due to the curvature of the critical hypersurface: If both scalar masses, which are relevant couplings, are chosen to deviate from their fixed-point values already above the Planck scale, gravitational fluctuations enforce a tiny non-vanishing value for the portal coupling. Its value is too small  to substantially impact the Higgs sector. }. Thus the scalar $\phi_d$ decouples at all scales.
As a result the Higgs mass stays unchanged at low energies.

To circumvent this decoupling result an additional degree of freedom needs to couple to $\phi_d$ such that its fluctuations generate a non-vanishing portal coupling above or below the Planck scale. Refs.~\cite{Reichert:2019car,Hamada:2020vnf} choose $\phi_d$ to be charged under an additional gauge symmetry. 
The corresponding interactions regenerate the portal coupling below the Planck scale, similar to what happens for the SM Higgs at the fixed point with vanishing but relevant Yukawa and gauge coupling, see above.

As an alternative \cite{Eichhorn:2020kca} introduces an additional dark fermion $\psi_d$ with a Yukawa coupling to the dark scalar that generates an interacting fixed point for the portal. 
We follow \cite{Eichhorn:2020kca} and hence consider the scale-dependent effective action
\be
  \Gamma_k = \Gamma_k^\text{visible} + \Gamma_k^\text{dark} + \int \dd[4]{x} \sqrt{g} \left(\frac{1}{16\pi \bar{G}_N} (2 \bar{\Lambda} - R) + \frac{\lambda_\text{HP}}{4} \phi_v^2 \phi_d^2 \right),\label{eq:Gammadark}
\ee
with
\bea
  \Gamma_k^\text{visible} = \int \dd[4]{x} \sqrt{g} &\Big(& \frac{Z_\phi}{2} g^{\mu\nu} \partial_\mu \phi_v \partial_\nu \phi_v - \xi_v \phi_v^2 R + \frac{\bar{m}_v}{2} \phi_v^2 \nonumber \\
   &+& \frac{\lambda_v}{8} \phi_v^4 + i Z_{\psi} \bar{\psi}_v \slashed{\nabla} \psi_v + i y_v \phi_v \bar{\psi}_v \psi_v \Big),\label{eq:Gammavisible}
\eea
and $\Gamma_k^\text{dark}$ follows by the replacement $v \rightarrow d$ in \eqref{eq:Gammavisible}.
Here $R$ is the Ricci scalar, $\bar{\Lambda}$ is the cosmological constant and $\bar{G}_N$ is the Newton coupling. The coupling $\lambda_v$ is the representative of the quartic Higgs coupling $\lambda_\text{H}$ in the SM.\\
This opens up a new universality class at which the dark and visible Yukawa couplings, scalar masses and non-minimal couplings are non-vanishing and in turn generate a non-vanishing portal coupling. At the same time, all couplings except for the Newton coupling, cosmological constant and the two mass parameters are irrelevant, endowing this fixed point with high predictive power. Here, we will investigate the impact of this extended dark sector on lower bounds on the Higgs mass. Our aim is to discover whether the single free parameter in the dark sector, namely the dark scalar mass, enables a lowering of the lower bound on the Higgs mass in an asymptotically safe setting.

\section{Higgs mass in an asymptotically safe dark portal model}
\label{sec:higgs_mass_extended}
We investigate whether the presence of a dark scalar can lower the predicted value of the Higgs mass in asymptotic safety, bringing it into agreement with observations. In order to generate a non-vanishing Higgs portal coupling, we consider an extended dark sector with a dark scalar and a dark fermion. We work in a toy model for the SM, with a real scalar for the Higgs and a Dirac fermion for the top quark. The flowing action we consider is given by Eq.~\eqref{eq:Gammadark} and \eqref{eq:Gammavisible}.
The beta functions can be found in Ref.~\cite{Gies:2010mqh} for the simple Yukawa model and \cite{Eichhorn:2020sbo} for the full gravity-matter system specified in Eq.~\eqref{eq:Gammadark} and \eqref{eq:Gammavisible}. 

We treat the fixed-point values of $G_{\ast}$ and $\Lambda_{\ast}$ as free parameters in this setting, in order to understand the phenomenological constraints on the gravitational parameter space. We limit ourselves to the regime $\Lambda_{\ast}< -3.3$, that might be reached in the presence of SM matter \cite{Dona:2013qba}. In this region, $f_y>0$ holds in the beta function for the Yukawa couplings $y_{v(d)}$ \cite{Eichhorn:2016esv,Eichhorn:2017eht,Eichhorn:2017ylw}
\be
  \beta_{y_{v(d)}} = \frac{5}{16\pi^2} y_{v(d)}^3 - f_y y_{v(d)}\label{eq:betay}
,\ee
where $f_y$ encodes the effect of gravitational fluctuations and depends on the gravitational fixed point couplings $\Lambda_\ast$ and $G_\ast$\footnote{There also is a dependence on the non-minimal coupling $\xi_{v(d)}$ and the mass that we neglect in our discussion but consider in our numerical results, see Ref.~\cite{Eichhorn:2020sbo} for details.}. 
At negative $f_y$, Eq.~\eqref{eq:betay} results in an upper bound on the IR value of the Yukawa which lies at zero. This is incompatible with a measured non-vanishing top-Yukawa coupling \cite{CMS:2019art,Agaras:2020zvy}. At positive $f_y$, the upper bound is shifted to non-zero values \cite{Eichhorn:2017ylw}, because an IR attractive interacting fixed point is present in Eq.~\eqref{eq:betay}.

The beta functions for both Yukawa couplings feature four fixed points: one at which both Yukawa couplings vanish, two at which one of the two Yukawa couplings has a non-vanishing value while the other one vanishes, and finally one at which both Yukawa couplings are non-vanishing. At this fully interacting fixed point, the Yukawa interactions break shift symmetry in both scalars. As a consequence, all scalar interactions are induced. This in particular includes the portal coupling. We will focus on this fixed point in the following. 

The fixed point in question is highly predictive \cite{Eichhorn:2020kca,Eichhorn:2020sbo}: The three quartics $\lambda_v, \lambda_d$ and $\lambda_\text{HP}$, the Yukawa couplings $y_v, y_d$ and the non-minimal couplings $\xi_v, \xi_d$ are irrelevant at this fixed point. In the matter sector, the two mass parameters $m_v^2$ and $m_d^2$ remain as the only relevant couplings. 

We start the RG flow to the IR at the fixed point. Deviations from quantum scale symmetry can occur along each of the two relevant directions, $m_{v,d}^2$.\\
To mimic the SM Higgs sector, we tune the UV deviation in $m_v^2$ from its fixed-point value such that $\phi_v$ undergoes spontaneous symmetry breaking along the RG flow and acquires a vacuum expectation value of $v_v \approx 246\, \text{GeV}$ in the IR.\\
For $m_d^2$ we can adjust the initial conditions such that $\phi_d$ either stays in the symmetric phase or undergoes spontaneous symmetry breaking. If $\phi_d$ remained in the symmetric phase, then the massive scalar $\phi_d$ could decay into the massless fermion $\psi_d$. This might lead to an over-abundance of relativistic degrees of freedom. We hence focus on the case where $\phi_d$ undergoes spontaneous symmetry breaking such that the dark fermion acquires a mass. We work in terms of the reparameterized potential \eqref{eq:potential}.
In this potential, fluctuations around the minimum $\langle \phi_{v(d)} \rangle = v_{v(d)}$ have masses $M_{v/d}$ given by Eq.~\eqref{eq:masses}.
The value of $M_d^2$ in the IR is adjusted in terms of the relevant perturbation in $m_d^2$ in the UV. The visible mass $M_v$ can be computed, once $v_v \approx 246\,\text{GeV}$ is used to fix the second relevant parameter, $m_v^2$. The remaining couplings are all canonically marginal and irrelevant at the fixed point. They are thus fixed at all scales as a function of $M_d$.\\

The reference point for the visible mass in our toy model is different from the Higgs mass in the SM due to the absence of gauge field fluctuations. 
 In the SM, gauge field fluctuations lead to an increase of the top Yukawa coupling towards the IR. In our toy model, the Yukawa coupling instead decreases towards the IR. The smaller Yukawa coupling causes a less strong increase of the Higgs quartic coupling towards the IR, leading to a lower Higgs mass.
 For fixed point values $\Lambda_\ast = -6.52$ and $G_\ast$ adjusted such that $y_\ast = 0.37$ and without a dark sector we obtain $M_v \approx 73\, \text{GeV}$.
Similarly, the quantitative amount of shifts in the visible mass that can be achieved due to the dark sector may differ. We expect that the qualitative (and semi-quantitative) effects of the dark sector remain the same. They will be our main focus in the following.\\

\begin{figure}[!t]
\centering
\includegraphics[width=0.7\linewidth]{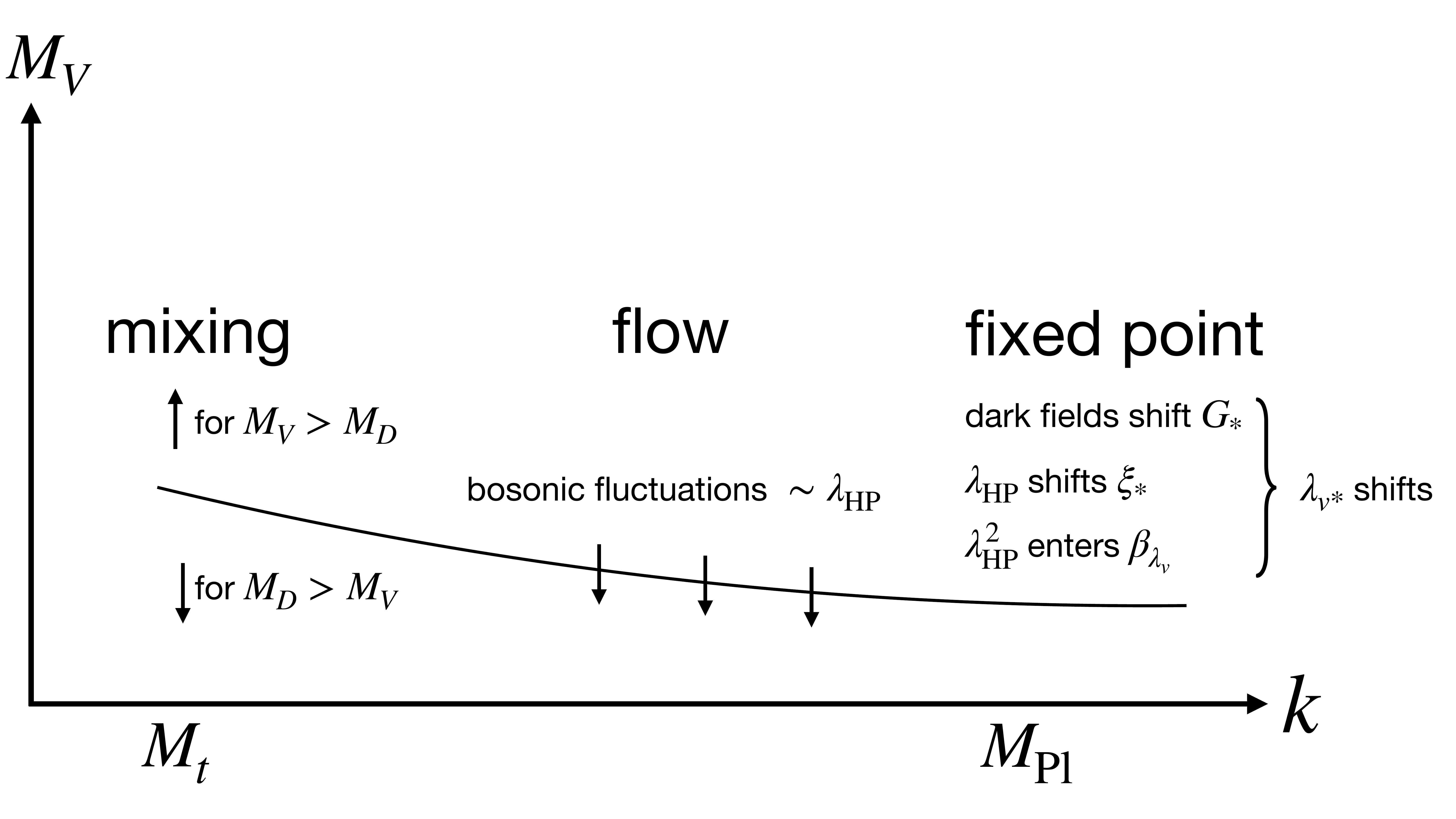}
\caption{\label{fig:sketch} Schematic depiction of all effects on the visible mass in asymptotic safety.}
\end{figure}

The dark sector impacts the visible mass $M_v$ through the non-vanishing portal coupling  in five ways.  Three of these (UV1, UV2 and UV3) are shifts in the UV initial conditions for $\lambda_v$. The fourth (F) is a change in the flow of $\lambda_v$ to the IR. The fifth (IR) is a mixing effect in the IR, see Fig.~\ref{fig:sketch} for a schematic depiction of the effects.

\begin{figure}[!t]
\centering
\includegraphics[width=0.5\linewidth]{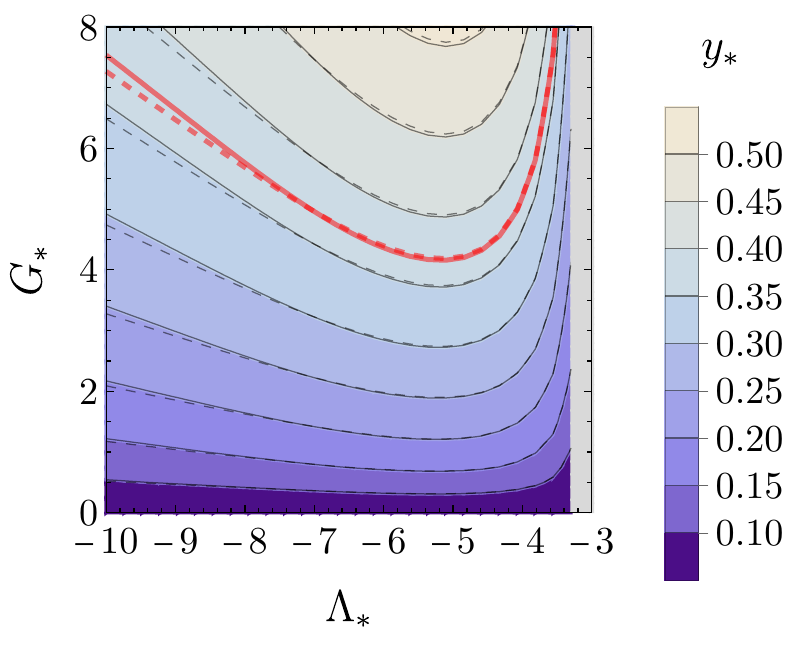}
\caption{\label{fig:Yukawacontours} 
We show contours of constant fixed-point value for the Yukawa coupling in the $G_{\ast}-\Lambda_{\ast}$-plane. Contours without a dark sector are shown in dashed, contours with the dark sector in continuous lines. The contours at $y_{\ast}=0.37$, which determine $G_{\ast}(\Lambda_{\ast})$ for our \emph{fixed-Yukawa scan} are shown in red lines.
}
\end{figure}

\begin{enumerate}
  \item[(UV1)] The gravitational fixed-point values are shifted due to quantum fluctuations of all dark sector fields. 
    In turn, this leads to shifts in the fixed-point values in the matter sector. 
    Thus, the UV initial conditions for the Higgs quartic coupling with and without dark sector would shift, 
    even if the dark sector would be completely decoupled from the visible sector.
  \item[(UV2)] The portal coupling contributes directly to $\beta_{\lambda_v}$ with a term $\sim \lambda_\text{HP}^2$.
    This shifts $\lambda_{v\, \ast}$ to smaller values than in the absence of a portal.
  \item[(UV3)] The portal coupling contributes indirectly to $\beta_{\lambda_v}$: The non-minimal coupling, which enters $\beta_{\lambda_v}$ linearly, 
    depends on $\lambda_{\rm HP}$ linearly. Thus $\lambda_{v\,\ast}$ can be shifted to larger or smaller values, depending on the sign of $\lambda_{\rm HP\, \ast}$. \\
  The shift of $\lambda_{v,\ast}$ depends on the balance of all three UV effects and does not have a unique sign across the gravitational parameter space.
  \item[(F)] The $\lambda_\text{HP}^2$-term in $\beta_{\lambda_v}$ changes the flow of $\lambda_v$ to the IR. The integrated effect is negative, i.e., it
    decreases the quartic coupling in the IR. This decreases the Higgs mass in the absence of mixing. 
  \item[(IR)] The portal coupling causes mixing between $\phi_v$ and $\phi_d$. 
    Mixing increases (decreases) the resulting mass $M_V$ for the visible scalar for $M_V > M_D$ ($M_V<M_D$).
\end{enumerate}
In the following we will study each of these effects individually and investigate how they compare quantitatively.
We achieve this by performing two types of parameter scans
\begin{itemize}
  \item In a \emph{gravitational scan}, we vary the gravitational fixed-point values freely and explore the resulting fixed-point values in the matter sector.
  \item In a \emph{fixed-Yukawa scan} we vary $\Lambda_{\ast}$ in the region $\Lambda_{\ast}<-3.3$ and adjust $g_\ast$ such that $y_\ast = 0.37$ remains constant. These scans follow the contour marked in red in Fig.~\ref{fig:Yukawacontours}.
\end{itemize}
In addition, we distinguish the cases \emph{with} and \emph{without} a dark sector.\\

The study of (UV1) - (UV3) is specific to our setup. The effect on the flow (F) and the IR mixing (IR) has been discussed before in an effective field theory context, see, e.g., \cite{Lebedev:2012zw,Elias-Miro:2012eoi,Gonderinger:2009jp,Cline:2013gha,Khoze:2014xha}, see Sec.~\ref{sec:Higgsmassportal}. For a study employing the FRG in this context, see Ref.~\cite{Eichhorn:2015kea}.\\
We caution the reader that any quantitative comparison should only be viewed as a statement about the relative size of the different effects in our toy model. 
Any comparison to the measured Higgs mass will need to take into account a more elaborate visible sector than the one in our toy model.

\subsection{The UV regime}
  \begin{figure}[!t]
 \begin{tabular}{cc}
 \includegraphics[width=0.45\textwidth]{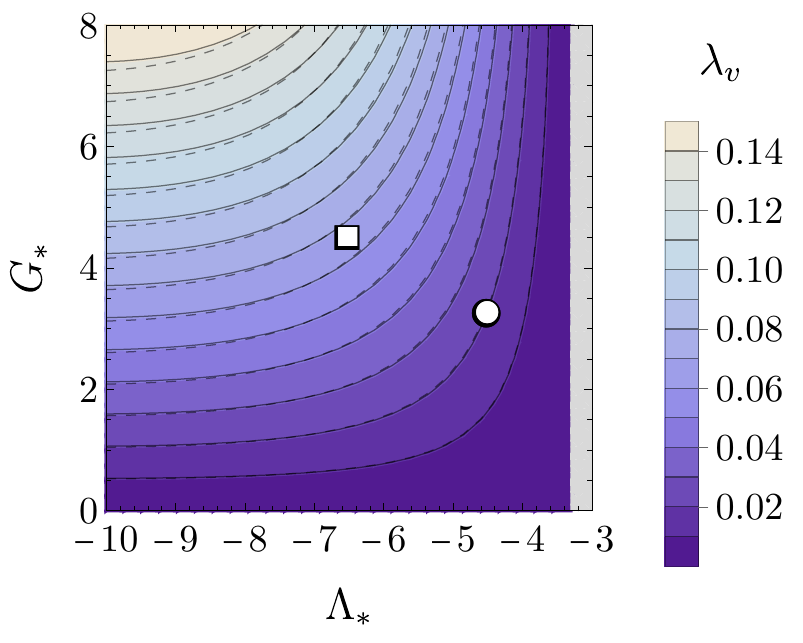} & \includegraphics[width=0.45\textwidth]{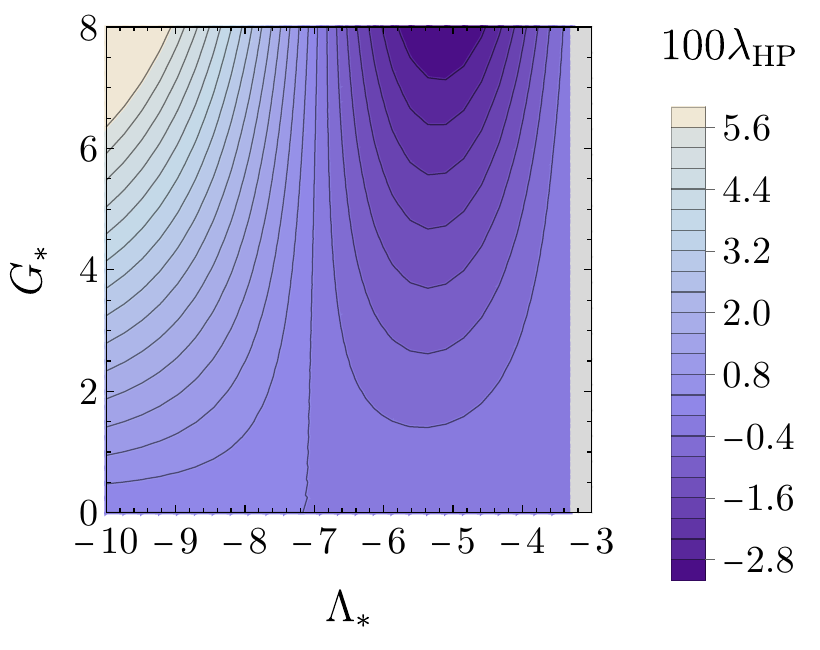}
 \end{tabular}
   \caption{\label{fig:quartic_contour}
    Contours for the fixed-point value $\lambda_{v \ast}$ ($\lambda_\text{HP}$) of the visible quartic coupling on the left (right) as a function of the gravitational fixed-point values $G_{\ast}$ and $\Lambda_\ast$ in a gravitational scan. The coupling $\lambda_v$ is the representative of the quartic Higgs coupling in our toy model. The dot (square) marks the position of the gravitational fixed point without (with) a dark sector in an approximation detailed in the main text. In the left plot solid (dashed) contours mark the value with (without) the contributions from a dark sector.
   }
 \end{figure}
 \begin{figure}
   \centering
   \setlength\tabcolsep{1.25em}
   \begin{tabular}{cc}
   \includegraphics[height=0.175\paperheight]{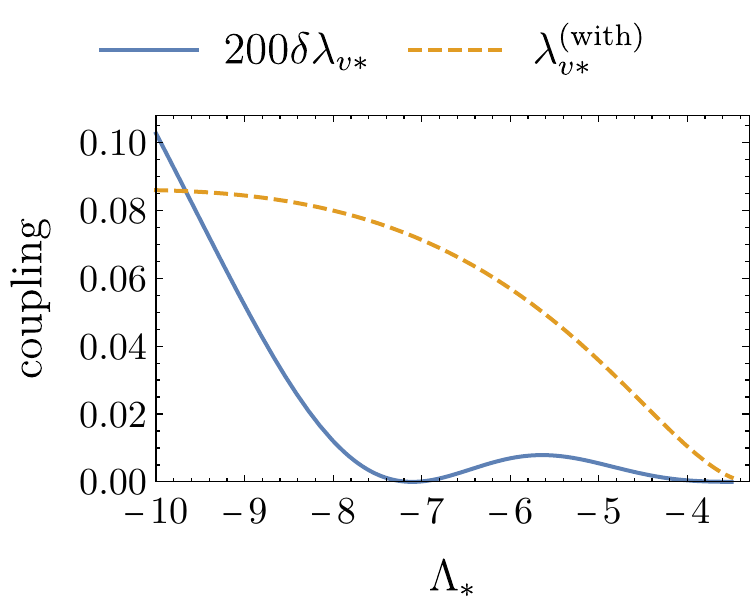} & \includegraphics[height=0.175\paperheight]{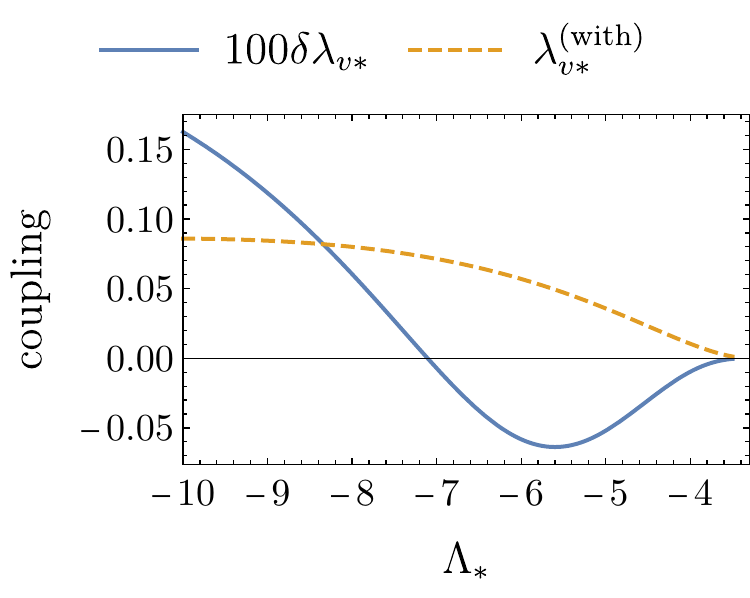} \\
    \qquad\quad(a) & \qquad\qquad(b)
   \end{tabular}
   \caption{\label{fig:delta_lambda}
    Difference $\delta \lambda_{v\ast} = \lambda_{v*}^\text{(without)} - \lambda_{v*}^\text{(with)}$ of the quartic coupling $\lambda_v$ without and with a dark sector for varying fixed-point values of the cosmological constant $\Lambda_\ast$ at fixed Newton coupling $G_{\ast} = 4.55$, i.e., in a gravitational scan. (a) $\lambda_{v*}^{(\text{without})}$ is computed by setting $\lambda_\text{HP} = 0$ in $\beta_{\lambda_v}$, while all other couplings are set to their $\lambda_\text{HP}$-dependent fixed-point values.  The direct contribution $\sim \lambda_\text{HP}^2$ lowers $\lambda_{v*}$, causing $\delta \lambda_{v\ast}>0$. (b) $\lambda_{v*}^{(\text{without})}$ is computed by solving the full matter beta functions self-consistently. Negative $\lambda_\text{HP}<0$ implies negative $\delta \lambda_{v\ast} < 0$.
   }
 \end{figure}
 \begin{figure}
   \centering
   \begin{tabular}{ccc}
     \includegraphics{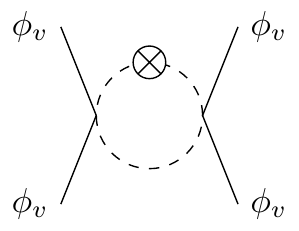} & \includegraphics{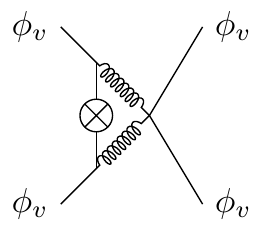} & \includegraphics{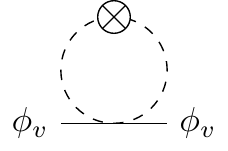} \\
     $\beta_{\lambda_v} \sim \lambda_\text{HP}^2$ & $\beta_{\lambda_v} \sim \xi_{}^3$ & $\beta_{\xi_v} \sim \lambda_\text{HP}$
   \end{tabular}
   \caption{\label{fig:diags}
   Diagrams representing different contributions of the portal coupling $\lambda_\text{HP}$ to the beta function of the visible quartic self-coupling $\lambda_v$. A solid line represents the visible scalar, a dashed line the dark scalar, and a wavy line the graviton propagator respectively.
    The cross vertex corresponds to a regulator insertion. For each of the diagrams, there are corresponding diagrams of the same structure with regulator insertions on one of the other internal lines. 
    (a) Direct contribution $\sim \lambda_\text{HP}^2$ from dark scalar loop. (b) Indirect contribution from graviton triangle which is odd in the non-minimal coupling $\xi$, whose beta function in turn has a contribution (c) odd in $\lambda_\text{HP}$ (the same remark applies, to the flow of squared scalar masses, but typically $m_{v,d\ast}^2 \ll \xi_{v,d\ast}$ by roughly an order of magnitude).
   }
 \end{figure}
 \begin{figure}
   \centering
   \includegraphics[width=0.45\textwidth]{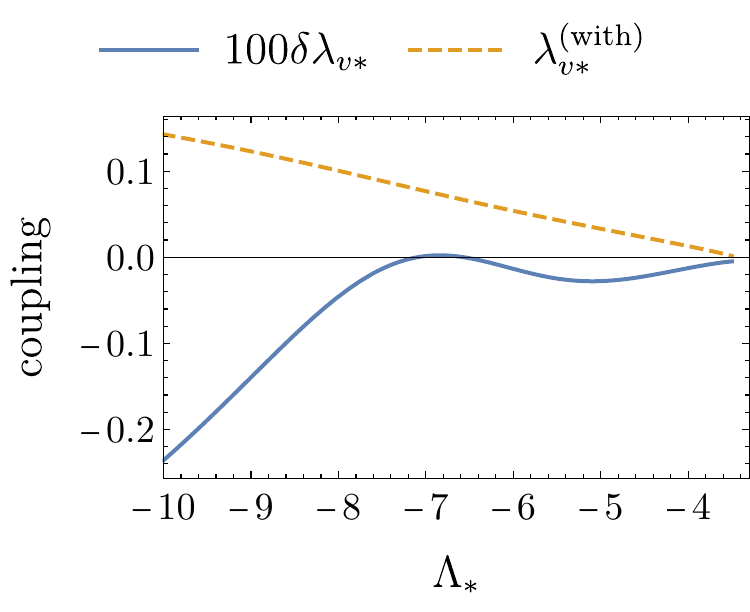}
   \caption{\label{fig:deltaLambdaUVofYuk} 
    Difference $\delta \lambda_{v\ast} = \lambda_{v*}^\text{(without)} - \lambda_{v*}^\text{(with)}$ of the quartic coupling without and with a dark sector for constant  fixed-point value of Yukawa coupling, i.e., in a fixed-Yukawa scan. Here $\lambda_{v*}^\text{(without)}$ is computed by solving the full matter beta functions self-consistently. For almost all values of $\Lambda_\ast$, the inclusion of the dark sector leads to larger values of $\lambda_{v*}$.
   }
 \end{figure}
 In our asymptotically safe toy model, the quartic couplings are predicted at the Planck scale. 
 The predicted values in turn  depend on the fixed-point values of the gravitational and other matter couplings. We discuss these dependencies separately.\\
 The gravitational fixed-point values depend on the number of matter fields of different spin \cite{Dona:2013qba}. 
 We compute them in the presence of SM matter with the beta functions reported in Ref.~\cite{Eichhorn:2017ylw} neglecting the back-reaction of non-vanishing masses and non-minimal couplings on the gravitational fixed point values. Due to the smallness of the fixed-point values in the matter sector, this is a viable approximation.
 In Fig.~\ref{fig:quartic_contour}, the fixed-point values $(G_{\ast}, \Lambda_{\ast})$ with and without a dark sector are indicated by a square and a dot, respectively. 
 The shift in $(G_{\ast}, \Lambda_{\ast})$ is nearly orthogonal to the contours of constant $\lambda_{v\, \ast}$. The fixed-point value of $\lambda_v$ nearly doubles because of this shift in $(G_{\ast}, \Lambda_{\ast})$.\\
 The (gravitational) fixed-point values have systematic uncertainties because the dynamics is truncated. Therefore, the square and dot in Fig.~\ref{fig:quartic_contour} should be understood as coming with significant uncertainties. These are difficult to estimate, but might even be as large as the difference in values between square and dot, see, e.g., \cite{Eichhorn:2017ylw} for an estimate.
 Therefore we treat $G_{\ast}$ and $\Lambda_{\ast}$ as free parameters in much of our investigation below.\\
 The fixed-point value of $\lambda_v$ depends on $\lambda_{\rm HP\, \ast}$ and therefore changes, when the dark sector is included, even when $G_{\ast}$ and $\Lambda_{\ast}$ are (artificially) held constant. This effect is quantitatively small, cf.~Fig.~\ref{fig:delta_lambda}, because
for most of the parameter space we explore, $\lambda_{v\ast} > \lambda_{\text{HP}\ast}$. This difference is generated by fermionic fluctuations, because only $\beta_{\lambda_v}$ contains a term $\sim y_v^4$, while $\beta_{\lambda_\text{HP}}$ contains no such term. Fermionic fluctuations induce a quartic self-coupling but not the portal coupling\footnote{ This statement is expected to hold at all loop orders/all orders of an FRG approximation, because the portal coupling breaks shift symmetry in the dark and the visible sector, whereas the Yukawa coupling only breaks shift symmetry in the visible sector. Thus the portal coupling is symmetry protected even at finite Yukawa coupling.}. Other contributions to $\beta_{\lambda_v}$ and $\beta_{\lambda_{\rm HP}}$ are subleading compared to this fermionic contribution. In turn, because of the small value of $\lambda_{\rm HP\, \ast}$, the addition of the dark sector results in a small change in $\lambda_{v\,\ast}$.
 This is exemplified by the contours of constant $\lambda_{v\ast}$ with and without portal coupling in the $(G_{\ast},\Lambda_{\ast})$ plane in Fig.~\ref{fig:quartic_contour}. The difference is typically not larger than a few percent. \\
 
 To further explore the difference we first perform a gravitational scan: The shift $\delta\lambda_{\ast} = \lambda_{v \ast}^{(\text{without})}- \lambda_{v \ast}^{(\text{with})}$ (at fixed gravitational coupling) arises due to both a direct contribution $\sim \lambda_\text{HP}^2$ in $\beta_{\lambda_v}$ as well as indirect contributions.\\
The direct contribution is even in $\lambda_{\rm HP}$ and thus lowers $\lambda_{v}$.
 To isolate the effect of this direct contribution (given diagrammatically by Fig.~\ref{fig:diags}(a)), we evaluate $\lambda_v^{(\text{without})}$ by switching off the portal contribution in $\beta_{\lambda_v}$. We determine all other couplings self-consistently at finite $\lambda_\text{HP}$ by solving the corresponding fixed-point conditions. The result is displayed in Fig.~\ref{fig:delta_lambda}(a). As expected, the direct contribution $\lambda_\text{HP}^2$ comes with a positive sign and lowers the visible quartic self-coupling $\lambda_{v\ast}$.\\
 The indirect contribution is odd in $\lambda_\text{HP}$ and thus lowers or increases $\lambda_{v \ast}$, depending on the sign of $\lambda_{\rm HP}$.
 This indirect contribution, (given diagrammatically in Fig.~\ref{fig:diags}(b)) turns $\delta \lambda_v$ negative, when $\lambda_{\rm HP}$ is negative, see Fig.~\ref{fig:delta_lambda}(b).\\
 
 To keep the IR fermion mass approximately constant, we also perform a fixed-Yukawa scan. In Fig.~\ref{fig:deltaLambdaUVofYuk} we compare the resulting fixed point values in the case with and without  dark sector. We keep $\Lambda_\ast$ constant and vary $G_\ast$ to obtain a constant Yukawa coupling.  For constant fixed-point values of the Yukawa coupling, the dark sector shifts $\lambda_{v\ast}$ towards larger values almost everywhere. For $\Lambda_\ast > -6.5$, the effect described above applies: the diagrams given in Fig.~\ref{fig:diags} lead to a larger $\lambda_{v\ast}$. For $\Lambda_\ast \ll -6.5$, the shift in $G_\ast$ in the presence of the dark sector (cf.~Fig.~\ref{fig:Yukawacontours}) also leads to larger values of $\lambda_{v\ast}$. 
 
 Due to the smallness of $\lambda_{\text{HP}\ast}$ compared to $\lambda_{v\ast}$ (as discussed above), the combined effect of direct and indirect contributions is typically of the order of 1\textperthousand. This is very much subleading compared to the change in $\lambda_{v\ast}$ that occurs due to the shift in the gravitational fixed-point values in our truncation.

 \subsection{Flow towards the IR}
\begin{figure}
 \centering
 \includegraphics[width=.45\textwidth]{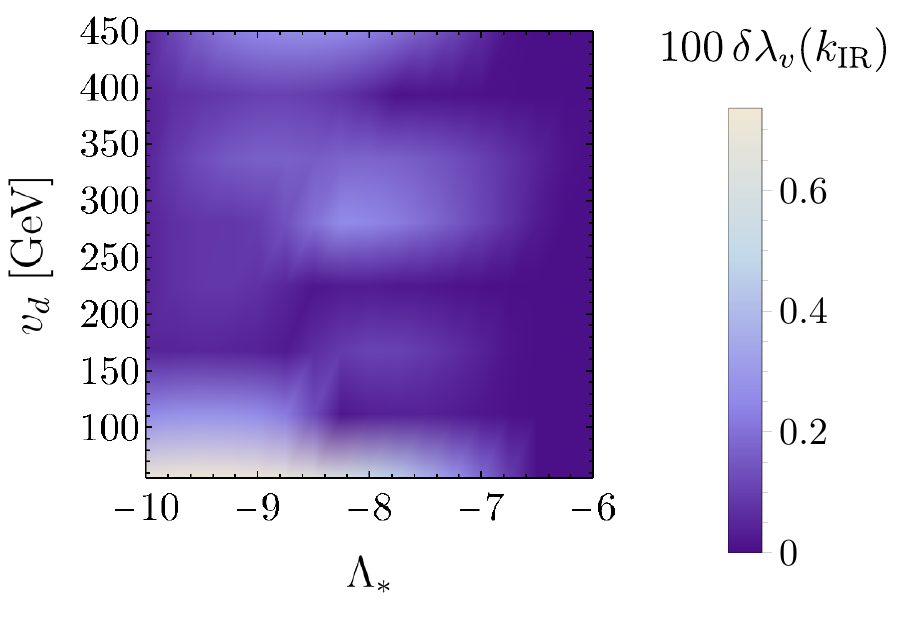}
   \caption{\label{fig:delta_mass_of_uv_portal}
    Difference $\delta\lambda_v(k_\text{IR}) = \lambda_v^{(\text{without})}(k_\text{IR}) - \lambda_v^{(\text{with})}(k_\text{IR})$ in the IR visible self-coupling between a setup that considers portal terms in the flow of $\lambda_v$ and a setup that ignores such terms. The UV initial conditions for both cases are the same, and obtained in a fixed-Yukawa scan.
   }
 \end{figure}

  Starting from the fixed point, we flow towards the IR. The flow of the quartic coupling receives an integrated negative contribution from the portal coupling. This lowers the quartic coupling in the IR.
  The effect occurs because bosonic fluctuations enter with a positive sign through a term $\sim \lambda_\text{HP}^2$ into the beta function for the quartic coupling. 
  This effect is small in our setup, cf.~Fig.~\ref{fig:delta_mass_of_uv_portal}, because $\lambda_\text{HP} \ll 1$.

 \subsection{Infrared masses}
\begin{figure}
   \centering
   \includegraphics[width=0.45\textwidth]{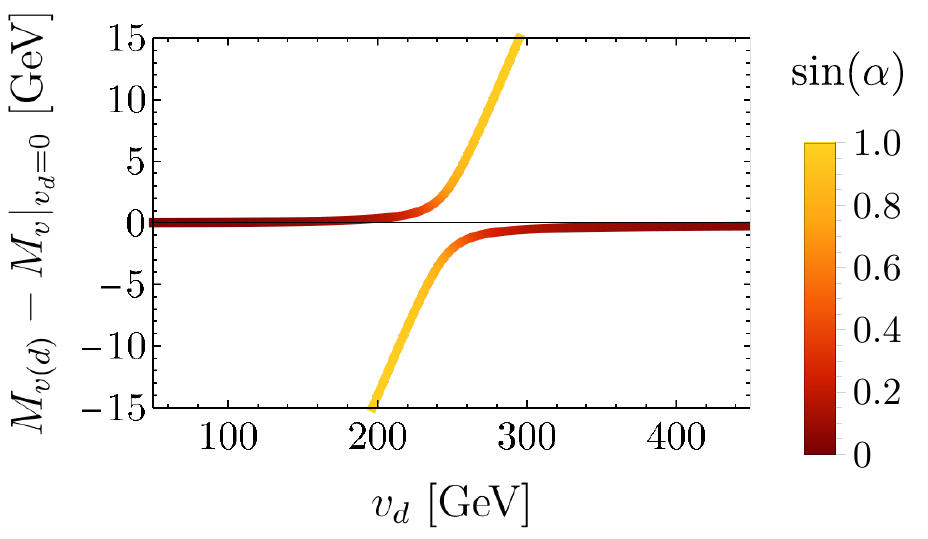}
   \caption{\label{fig:eigenvalues_mass_matrix}   
    Eigenvalues of the mass matrix as a function of the dark scalar vacuum expectation value for a fiducial set of IR quartic couplings $\lambda_v = \lambda_d = 8.79\times 10^{-2}$, $\lambda_\text{HP} = -6.22\times 10^{-3}$.  We choose these values as examples to illustrate the mixing between the two scalars. The overlap of the mass eigenstate with the dark scalar is colour coded. (The sine of the mixing angle by convention corresponds to the overlap of the heavier mass eigenstate with the dark scalar.)
   }
 \end{figure}

 In the IR one of two distinct scenarios is realized: either the dark scalar $\phi_d$ undergoes spontaneous symmetry breaking (SSB) and develops a vacuum expectation value or it maintains its $\mathbb{Z}_2$ symmetry. In the case of unbroken symmetry, the dark sector only slightly affects the visible mass. In the case of SSB, the dark sector can strongly affect the visible mass because the mass matrix becomes non-diagonal. Both its eigenvalues depend on $\lambda_{\rm HP}$ and are therefore shifted compared to the case of unbroken symmetry.
This shift in the eigenvalues of the mass matrix is illustrated in Fig.~\ref{fig:eigenvalues_mass_matrix}. The masses repel each other due to the non-vanishing portal coupling. This decreases the mass of the lighter scalar and increases the one of the heavier scalar.

 When the visible scalar is the lighter one, $M_v < M_d$, its mass is lowered further when the dark vev is lowered. For a shift in the visible mass of the order of $1\,\text{GeV}$, the corresponding mixing angle is of the order of $\sin(\alpha) \sim 0.3$.
 
 In consequence, the dark vev, which is a free parameter in asymptotic safety, can be used to dial in the desired value of the visible mass. Consequently, all properties of the dark sector are fixed. 

 \subsection{From the UV to the IR -- Contrasting EFTs and Asymptotic Safety}
 The visible mass in asymptotic safety is predicted as a function of the two vevs. The predicted value is determined by the five effects (three fixed-point shifts in the UV, change in the flow to the IR, mass mixing in the IR). We discuss the combination of the five effects. 
 For each of the three regimes (UV, flow to IR, IR), we contrast effective and asymptotically safe theories. In doing so we illustrate how asymptotic safety could strongly enhance the predictivity of a given theory. 

 In an effective perturbative field theory the quartic couplings in the UV are only constrained by the two inequalities 
\be
  \lambda_{v,d}>0, \qquad \lambda_\text{HP}^2 - \lambda_v \lambda_d >0\label{ineq:conds}
,\ee
ensuring a stable vacuum and by the requirement $\lambda_i/(16\pi^2)<1$ (or stricter perturbativity requirements). 
 Therefore, there are UV intervals of finite extent for all three couplings in an EFT. Each set of values $(\lambda_v, \lambda_d, \lambda_{\rm HP})$ 
 is translated to IR masses via the RG flow.
 One can adjust a combination of the portal coupling and the Higgs quartic coupling in the UV to obtain the correct IR Higgs mass even when the UV scale is chosen to be the Planck scale \cite{Lebedev:2012zw,Elias-Miro:2012eoi,Gonderinger:2009jp,Cline:2013gha,Khoze:2014xha}.
 
 The same freedom is absent in asymptotic safety where the quartic couplings are fixed uniquely because they are irrelevant couplings at an interacting fixed point.
 This provides unique initial conditions for the RG flow at the Planck scale.
 These in turn yield a highly constrained IR phenomenology. We compare the IR phenomenology to that of the EFT setting.
  
\begin{figure}
  \centering
  \includegraphics[scale=0.85]{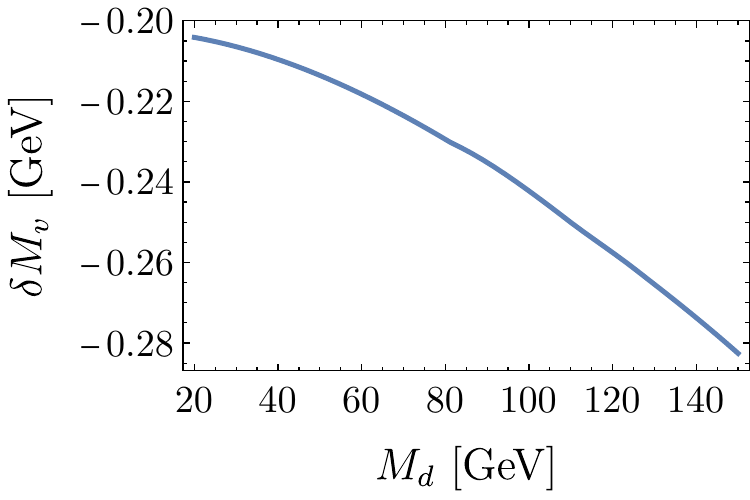}\qquad\includegraphics[scale=0.85]{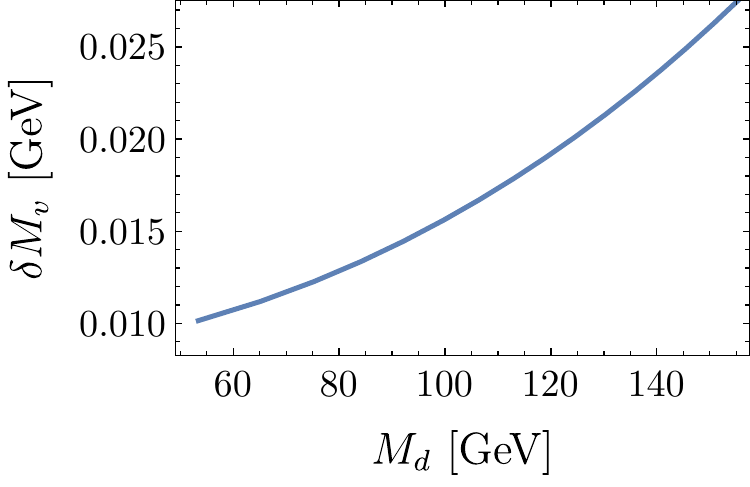}
   \caption{\label{fig:delta_mv_symmetic}
    Change of the mass $\delta M_v = M_v^{(\text{without})} - M_v^{(\text{with})}$ with and without a dark sector respectively in the case where the dark scalar does not undergo spontaneous symmetry breaking. The cosmological constant is $\Lambda_* = -10$ ($\Lambda_\ast = -6.52$) in the left (right) panel, while the Newton coupling $G_*$ is varied such that the Yukawa coupling is fixed at $y = y_* = 0.37$.
   }
 \end{figure}

 First, we consider the case in which the dark scalar does not undergo spontaneous symmetry breaking. 
 As apparent from Fig.~\ref{fig:delta_mv_symmetic}, the modifications are tiny. The dark sector does not change the Higgs mass substantially. This is a direct consequence of the magnitude of the portal coupling, $\lambda_\text{HP} \sim 10^{-3}$.
 The small value follows from the requirement of an asymptotically safe UV completion of our toy model. 
 Assuming that our results carry over to the full SM setting this would exclude large modifications to the Higgs mass due to the dark scalar without spontaneous symmetry breaking. This is different to the EFT case, where a sizeable portal coupling can be chosen (without violating the inequality \eqref{ineq:conds}).

\begin{figure}[!t]
  \centering
  \includegraphics[height=.125\paperheight]{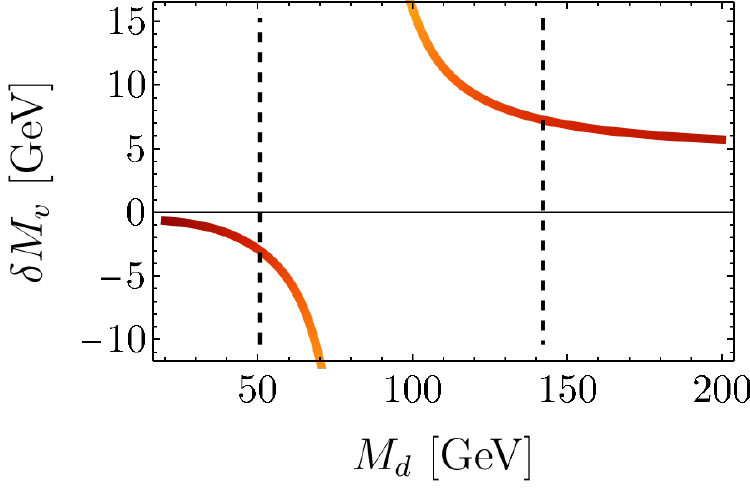}\qquad\includegraphics[height=.125\paperheight]{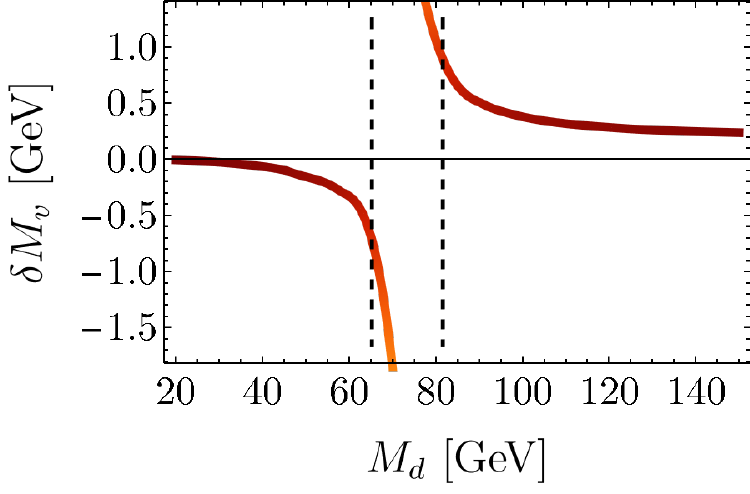}\quad \includegraphics[scale=.75]{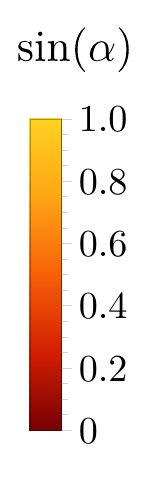}
   \caption{\label{fig:delta_mv_ssb}
    Change of the mass $\delta M_v = M_v^{(\text{without})} - M_v^{(\text{with})}$ with and without a dark sector in the case where the dark scalar does undergo spontaneous symmetry breaking. The cosmological constant is set to $\Lambda_\ast = -10$ (left) and $\Lambda_\ast = -6.52$ (right), while the Newton constant is adjusted so that the visible Yukawa coupling remains fixed at $y = y_\ast = 0.37$, i.e., we perform a fixed-Yukawa scan. The mixing angle is colour-coded; the dashed line demarcates where $\sin(\alpha) < 0.3$, which is approximately the experimental constraint on the mixing of the SM Higgs.
   }
 \end{figure}

 Second, we consider the case where $\phi_d$ undergoes spontaneous symmetry breaking and acquires a vacuum expectation value. As apparent from Fig.~\ref{fig:delta_mv_ssb} the modifications of the visible mass can be sizeable. Out of the various effects altering the mass, the IR mixing effect is dominant. For $v_d > v_v$ this allows to lower the visible mass $M_v$.
 A lowering of about $\sim 7\,\text{GeV}$ ($\sim 1\,\text{GeV}$) implies a mixing angle $\sin(\alpha)\sim 0.3$ for $\Lambda_\ast = -10$ ($\Lambda_\ast = -6.5$). \\
 We caution that these numbers are obtained in our toy model, but expect that similar-sized effects can be achieved if an extension of our toy model to the full SM is asymptotically safe.

 Assuming that an extension to the full SM exists, we make the following observations:\\
 The Higgs mass might be lower in the presence of a portal coupling to a dark scalar. This could reconcile the Higgs mass predicted from asymptotic safety with observations\footnote{We assume a top mass of $173\,\text{GeV}$ here. If the top quark is lighter, the scenario from \cite{Shaposhnikov:2009pv} could be viable without a dark sector. If the top quark weighs about $173\,\text{GeV}$, our mechanism is applicable for the universality class at which some of the SM couplings are nonzero at the fixed point, but not in the scenario of \cite{Shaposhnikov:2009pv}, where all SM couplings  are zero at the fixed point, resulting in a vanishing portal coupling at all scales.}. \\
 Our results indicate that to achieve a sufficiently large impact on the Higgs mass, the dark scalar needs to undergo spontaneous symmetry breaking and should be heavier than the Higgs particle.\\
 A sizeable portal coupling requires the presence of additional dark degrees of freedom beyond the dark scalar. We focus on a dark fermion, because a dark gauge field may not yield a symmetric phase for the dark scalar in the UV, cf.~\cite{Eichhorn:2019dhg,Wetterich:2019rsn}.
 The  dark fermion $\psi_d$ becomes massive once the dark scalar undergoes SSB.
 In the IR, no relativistic degree of freedom beyond those of the SM remain.
 Interestingly, this is compatible with bounds on additional relativistic degrees of freedom that arise from Big Bang Nucleosynthesis \cite{Cyburt:2015mya,Pitrou:2018cgg}.
 
 Similar models have been studied in an effective field theory context as dark matter candidates \cite{Esch:2013rta,Bagherian:2014iia,Krnjaic:2015mbs}.
 Asymptotic safety constrains the parameter space of these models and is therefore not guaranteed to result in a viable dark-matter phenomenology.
 According to the preliminary study in \cite{Eichhorn:2020kca}, the dark fermion might be available as a viable dark matter candidate. Taken together with the results in the present paper, this strongly motivates a study of the SM together with a dark sector as in this paper.

\section{Conclusions}
\label{sec:conclusion}
There are promising indications that asymptotically safe quantum gravity could enhance the predictive power within the SM and some of its extensions. This could allow to compute the Higgs mass in (extensions of) the SM from first principles.\\
 Here, we have focused on a toy model that features a real scalar and a fermion as the SM Higgs and top respectively. We postulate a dark sector containing a dark scalar and a dark fermion coupled through a Yukawa coupling to each other and a portal coupling to the SM. In addition to lowering the Higgs mass compared to the pure SM case, the dark sector might simultaneously provide a dark-matter candidate.

The dark sector, which contains five canonically marginal or relevant couplings (i.e., five free parameters according to canonical power counting) contains a single relevant coupling at an asymptotically safe fixed point found within the approximation of \cite{Eichhorn:2020kca}. This free parameter can be used to vary the Higgs mass. We found that unless the dark scalar undergoes spontaneous symmetry breaking, the resulting modifications of the Higgs mass are small.
If the dark scalar undergoes spontaneous symmetry breaking, then the resulting modifications can become sizeable. The most relevant effect is the tree-level mixing between the dark and the visible scalar. The various UV effects related to the dark sector are relatively small. 

Once the single free parameter is used to obtain the measured Higgs mass, there is no free parameter left in the dark sector. Accordingly, the dark relic density and the cross-sections relevant for direct dark-matter searches could be predicted from first principles. If the indications from our toy model and approximation hold up, asymptotic safety is quite distinct from other approaches to BSM physics  with less predictive power and thus more freedom to match observational data.
For instance, observational data from cosmology constrain the dark sector. First, the dark fermion may not be massless in order to not violate constraints on the number of relativistic degrees of freedom at Big Bang Nucleosynthesis. Second, no field in the dark sector may be overproduced, so as to not overclose the universe. We find tentative indications that both constraints might be met: The dark fermion becomes massive, once the dark scalar acquires a vev, such that BBN constraints hold. The dark scalar is not stable, and therefore its relic density vanishes. The dark fermion is stable and has a finite relic density, which, according to estimates in \cite{Eichhorn:2020kca} is close to critical density.

 Overall, our results show how the predictive power of asymptotic safety could constrain the vast parameter space of BSM models. Combining these theoretical constrains with phenomenological constraints from particle physics and cosmology turns out to be nontrivial, as we show here. One may therefore hope that the asymptotic-safety paradigm results in a small set of viable models which make definite predictions for future experiments and observations.

\acknowledgments
A.~E.~is supported by a research grant (29405) from VILLUM FONDEN. M.~P.~is  supported by  a  scholarship  of  the  German  Academic  Scholarship Foundation. S.~R.\ acknowledges support by the Deutsche Forschungsgemeinschaft (DFG) through SFB 1143 (Project A07, Project id No.\ 247310070), the Würzburg-Dresden Cluster of Excellence ct.qmat (EXC2147, Project id No. 390858490), and the Emmy Noether programme (JA2306/4-1, Project id No. 411750675). M.~P.\ and S.~R.\ are grateful to CP3-Origins at the University of Southern Denmark for extended hospitality during various stages of this work.

\bibliography{references.bib}
\end{document}